\documentclass[fleqn,usenatbib]{mnras}

\usepackage{newtxtext,newtxmath}

\usepackage[T1]{fontenc}
\usepackage{ae,aecompl}
\usepackage{soul}

\usepackage{graphicx}	
\usepackage{amsmath}	

\usepackage{amssymb}	
\usepackage{color}
\usepackage{soul}

\newcommand{\gppr}{\stackrel{>}{\scriptstyle \sim}}
\newcommand{\gappr}{\raisebox{-0.4ex}{$\gppr$}}

\newcommand{\G}{{\it Gaia}}
\newcommand{\Msun}{M$_{\odot}$}

\usepackage[T1]{fontenc}
\usepackage{ae,aecompl}

\usepackage{newtxtext,newtxmath}

\title[The {\it Gaia} resolved white dwarf binary population]{A population synthesis fitting of the {\it Gaia} resolved white dwarf binary population within 100 pc}

\author[S. Torres et al.]{
S. Torres$^{1,2}$\thanks{E-mail: santiago.torres@upc.es}, P. Canals,$^{1}$, F. M. Jim\'enez-Esteban$^{3,4}$, 
\newauthor 
A. Rebassa-Mansergas$^{1,2}$, E. Solano$^{3,4}$  
\\
$^{1}$ Departament de F\'{\i}sica, Universitat Polit\`{e}cnica de Catalunya, c/Esteve Terrades 5, 08860 Castelldefels, Spain\\
$^{2}$ Institut d'Estudis Espacials de Catalunya, Ed. Nexus-201, c/Gran Capit\`a 2-4, 08034 Barcelona, Spain\\
$^{3}$Departmento de Astrof\'{\i}sica, Centro de Astrobiolog\'{\i}a (CSIC-INTA), ESAC Campus, Camino Bajo del Castillo s/n,\\
E-28692 Villanueva de la Ca\~nada, Madrid, Spain\\
$^{4}$Spanish Virtual Observatory, Spain\\
}

\date{Accepted XXX. Received YYY; in original form ZZZ}

\pubyear{}
\hypersetup{draft}
\begin{document}
\label{firstpage}
\pagerange{\pageref{firstpage}--\pageref{lastpage}}
\maketitle

\begin{abstract}
The {\it Gaia} mission has provided an unprecedented wealth of information about the white dwarf population of our Galaxy. In particular, our studies show that the sample up to 100\,pc from the Sun can be considered as practically complete. This fact allows us to estimate a precise fraction of double-degenerate ($1.18\pm 0.10$\%) and white dwarf plus main-sequence stars ($6.31\pm0.23$\%) among all white dwarfs through comoving pairs identification. With the aid of a detailed population synthesis code we are able to reproduce synthetic white dwarf populations with nearly identical fractions as the observed ones, thus obtaining valuable information about the binary fraction, $f_{\rm b}$, initial mass ratio distribution, $n(q)$, and initial separation distribution, $f(a)$, among other parameters. Our best-fit model is achieved within a $1\sigma$ confidence level for $f(a)\propto a^{-1}$, $n(q)\propto q^{n_q}$, with $n_q=-1.13^{+0.12}_{-0.10}$ and $f_{\rm b}=0.32\pm 0.02$. The fraction of white dwarf mergers generated by this model is $9\sim16\%$, depending on the common-envelope treatment. As sub-products of our modelling we find that around $1\sim3\%$ of the white dwarf population are unresolved double-degenerates and that only $\sim1\%$ of all white dwarfs contain a He-core. Finally, only a mild kick during white dwarf formation seems to be necessary for fitting the observed sky separation of double-degenerate systems. 

\end{abstract}


\section{Introduction}
\label{s-intro}

White dwarfs are the most common remnant of single evolution of low- to intermediate-mass main-sequence stars and also a common product of binary evolution. Since white dwarfs are stellar objects in which fusion reactions have ceased, their evolution is ruled by a slow cooling process whose characteristics are reasonably well understood from a theoretical point of view \citep{Althaus2010}. Supported by the degeneracy pressure of electrons that prevent gravitational collapse, the oldest white dwarfs can have cooling ages as long as  $10\,$Gyr or even more. Therefore, the analysis of these objects can provide us with reliable information about the past history and evolution of our Galaxy, as well as the stellar evolutionary processes that take place \citep[e.g][]{GBerro2016}. For instance, white dwarfs have been used to constrain the halo dark matter content in our Galaxy \citep[e.g.][]{Flynn2003,Torres2008}, the recycling of material to the interstellar medium \citep[e.g.][]{Barstow2006}, and the central objects in the generation of type Ia supernovae through the different possible scenarios \citep[e.g.][]{Livio2018}.

However, white dwarfs are intrinsically faint objects, a fact that prevents the construction of complete samples far beyond the solar neighborhood. Initial attempts were restricted to a few hundred objects and up to 20\,pc and 25\,pc from the Sun \citep{Holberg2008, Holberg2016} or up to 40\,pc \citep{Limoges2013} but this last case restricted to the northern celestial hemisphere. On the other hand, it was not until the arrival of large-scale automated surveys such as the Sloan Digital Sky Survey \citep{York2000}, the Pan-STARRS collaboration \citep{Kaiser2002}, the SuperCosmos Sky Survey \citep{Hambly1998} or the {\it Gaia} mission \citep{GaiaDR12016}, among other examples, that the number of known white dwarfs has exponentially increased. Especially, the {\it Gaia} mission has provided us with an extraordinary wealth of information. In its last third early data release (EDR3), the analysis of {\it Gaia} data reveals the existence of around $359,000\,$ white dwarf candidates \citep{Fusillo2021} and a million of binary systems \citep{ElBadry2021} of which $\simeq$16,000 are identified as white dwarf plus main sequence binaries (WDMS) and $\simeq$1,400 as double degenerate systems (DWD). 

This huge amount of information allows the construction of statistically significant samples that can be used to test the different theoretical models. Of particular interest is the building of an unbiased and complete census of the different sub-populations of white dwarfs, i.e., those coming from single, binary or merger evolution. As it is well known, the white dwarf population coming from single star evolution can be reasonably well modeled provided a relatively restricted set of parameters such as the age of the sample, the star formation rate, the initial mass function and the initial-to-final-mass relationship, among the most important ones \citep[e.g.][]{GBerro1999,Rowell2013}. However, in the case of binary evolution, the complexity of the physics involved increases the number of unknown parameters: a binary fraction, an initial separation distribution, an initial mass ratio distribution, a mass loss rate, a common-envelope treatment and several others \citep[e.g.][]{Hurley2002}. Moreover, many of these parameters are poorly constrained or in the best of the cases only partial estimates are derived from the analysis of one of the white dwarf sub-populations. In this sense, several population synthesis studies have been devoted to, for instance, close or unresolved white dwarf plus main sequence stars, \citep{Willems2004,Davis2010,Camacho2014,Cojocaru2017}, double degenerate white dwarfs \citep{Nelemans2001, Rebassa-Mansergas2020} or accreting white dwarfs \citep{Chen2014,Chen2015}. But only a few of these studies have approached the problem from a comprehensive perspective \citep{Toonen2017,ElBadry2018}. 

In particular, \citet{Toonen2017} thoroughly analyzed a sample of $\sim 100\,$white dwarfs within $20\,$pc from the Sun. The synthetic modeling of that sample is in reasonable agreement with the observed fraction of the different sub-populations except for the 
space density of resolved DWD. However, the relative low number of objects in the sample does not allow further statistical conclusions to be drawn. On the other hand, \cite{ElBadry2018} studied a larger sample containing nearly 3,000 WDMS and 400 DWD up to $200\,$pc. In their analysis they found that the observed separation distributions of WDMS and DWD systems have different breaks at $\sim 3,000\,$AU and $\sim 1,500\,$AU, respectively, arguing that a possible white dwarf kick occurs during its formation due to an asymmetric mass loss. However, a more detailed analysis of the initial binary parameters is deserved to test this hypothesis or prove that, on the contrary, the observed effect is the product of selection biases in the construction of the observed sample. 

In this paper we aim to analyze from a holistic perspective the single and binary population of white dwarfs derived from recent EDR3 {\it Gaia} data. We select the $100\,$pc sample as the best compromise between a nearly complete sample and a statistically significant sample \citep{Jimenez-Esteban18}. The selected sample contains nearly 13,000 systems and it allows us to derive through the identification of comoving pairs a precise fraction of the different sub-populations of white dwarfs. With the aid of a detailed population synthesis simulator which has been widely used on the study of the single and binary white dwarf population \citep[e.g.][]{GBerro1999,Torres2005, Camacho2014,Cojocaru2017}, we analyze the effects of the input parameters aimed to find the model that best fits the observed fractions of the different sub-populations, in particular the fraction of resolved WDMS and DWD systems.

The present paper is organized as follows. In Section 2 we detail the physical ingredients of our population synthesis code. In Section 3 we present our search methodology and the identification of comoving pairs extracted from {\it Gaia} EDR3 data. Section 4 is devoted to the analysis of the space parameter, firstly identifying those parameters who play a key role and, secondly, through a fine tuning analysis determining the best fit model. Finally, we summarize our results and present our conclusions.


\section{The synthetic binary white dwarf population}
\label{s:synt}
Our population synthesis code has been widely used in the study of the single \citep[e.g.][]{GBerro1999,Torres2005,Torres2016,Jimenez-Esteban18} and binary \citep[e.g.][]{Camacho2014,Cojocaru2017,Canals2018} white dwarf population, as well as on the disentanglement of the components of the Galaxy \citep[e.g.][]{Torres2019}, studies of open and globular clusters \citep[e.g][]{GBerro2010,Torres2015} and the Galactic bulge \citep{Torres2018}, among other applications. The code, based on Monte Carlo techniques, generates in a self-consistent way a population of single and binary stars for the three components of the Galaxy: thin and thick disk and stellar halo. A detailed description of the physical inputs used in the single population can be found in \cite{Jimenez-Esteban18}, while for the binary population we point the reader to \cite{Canals2018} and references therein. For the sake of conciseness, we briefly present here the main ingredients of both, single and binary, populations. 

First of all, by means of the binary fraction parameter, $f_b$, we randomly decided if the system is going to be a single star or a binary system. Single star evolution proceeds as follows. Masses of single main-sequence stars are randomly derived in the range $0.08-50\,$\Msun from the initial mass function from \cite{Kroupa2001}. Basically, this distribution presents an average power-slope of $\alpha_{\rm IMF}=-2.3$ for the range of solar masses and a flatter slope for masses below $0.5\,$\Msun. Additionally, we extend that distribution according to \cite{Sollima2019} by changing to a positive slope for masses below $0.16\,$\Msun. Although the exact distribution in the subsolar regime of the initial mass function does not play any role in the single white dwarf population (e.g. a $0.5\,$\Msun\ main-sequence star will take more than the Hubble time to become a white dwarf), it plays a capital role in the binary population since the companions may have time to become a white dwarfs. Regarding the birth time, in both cases (single and binary stars) it is randomly chosen from a star formation history. In particular, for the thin disk population we adopt a constant star formation rate lasting 9.2\,Gyr and with a dispersion in metallicity around the solar metallicity value, $Z=0.014$, as applied in \cite{Tononi2019}. In the case of the thick disk population we chose a sub-solar, $Z=0.001$, metallicity star formation rate peaked at 10\,Gyr in the past and extended up to 12\,Gyr, while halo stars were born according to a 1 Gyr burst of constant star formation and $Z=0.0001$ metallicity happened 13.5\,Gyr in the past. The spatial distribution is generated according to a double exponential profile for the thin and thick disk populations, and according to a spherical isothermal distribution for halo stars. At this point we can evaluate which stars have become white dwarfs and follow, then, their evolution by means of an updated set of white dwarf evolutionary cooling sequences. We applied those provided by La Plata Group -- see \cite{Althaus2015} and \cite{Camisassa2017,Camisassa2019} and references therein. These sequences encompass the full range of metallicities and white dwarf masses for CO-core and ONe-core white dwarfs and represent a set of full evolutionary calculations of their progenitor stars, starting at the Zero Age Main Sequence (ZAMS), all the way through central hydrogen and helium burning, thermally-pulsing AGB and post-AGB phases. In all cases, they incorporate an updated and detailed prescription of energy processes in white dwarf evolution such as residual hydrogen shell burning, energy released by latent heat, phase separation of carbon and oxygen due to crystallization and neon diffusion, among other processes. 

On the other hand, if the system resulted to be in a binary we followed its evolution based on the binary stellar evolution (BSE) code by \cite{Hurley2002} with some updates introduced in \cite{Camacho2014} and \cite{Cojocaru2017}. The code incorporates a detailed prescription for all the essential aspects of binary evolution, such as mass transfer and accretion, wind loss, common-envelope evolution, circularization and synchronization of orbits, angular-momentum loss, magnetic braking and gravitational radiation emission, among other effects. The primary masses\footnote{The primary star is the more massive component in the binary.} of the binary systems are randomly chosen following an initial mass function in the same way as for single stars, but their ratio, $q=M_2/M_1$, where $M_1$ and $M_2$ are the masses of the primary and secondary star, respectively, follows a particular initial mass ratio distribution, $n(q)$ (see Section \ref{s:mod} for further details). The initial orbital separation of the binary system is determined by randomly generating the semi-major axis $a$ and the eccentricity $e$. Additionally, a random inclination $i$ according to a $\sin i$ distribution is generated for those binary systems. For the semi-major axis we explored different distribution (see Section \ref{s:mod}) in the range $3 \le a/R_{\odot} \le 11\cdot10^{6}$ while for the eccentricity $e$ of the binary we adopted a standard thermal distribution, $f(e)=2e$ with $0\le e<1$ \citep{Heggie1975}. Each of the stars of the binary system is then left to evolve individually until, if it is the case, an eventually mass transfer episode or a common envelope episode occurs. For this last case, different theoretical treatments have been proposed, hampered them by the observational difficulties to constrain any model. In our case, we chose as our reference model the standard approximation of the $\alpha$-formalism in which it is assumed energy conservation of orbital and envelope-binding energy \citep{Ibeb+Livio93, Webbink08}. The $\alpha$-formalism introduces a set of parameters such as the $\alpha_{\rm CE}$ parameter, which accounts for the efficiency of the system to convert orbital energy into energy to expel the envelope, the binding energy parameter $\lambda$ which represents the ratio between the approximate and the exact expression of the binding energy and that depends on the structure of the primary star, and the $\alpha_{\rm int}$ parameter by which the internal energy of the system can be converted into kinetic energy to expel the envelope \citep{Han1995}. Although there is no general agreement about the best choice for these parameters, we adopted a value of $\alpha_{\rm CE}=0.3$, no internal energy and a $\lambda$ variable model as deduced in the analysis of close WDMS from SDSS by \cite{Camacho2014}. Additionally, we also compare our results with the $\gamma$-formalism. Introduced by \citet{Nelemans2000}, it takes into account the angular momentum conservation as the main ingredient in the common-envelope phase. We use the  particular $\gamma\alpha$ prescription model introduced in \citet{Toonen2017} where we adopt the values $\gamma=1.75$ and $\alpha\lambda=2$. Finally, as a result of common envelope evolution or a collision, some binary systems may coalesce leading in some cases to the formation of a merged white dwarf. The treatment of these evolutionary channels has been also included in our simulations following the prescriptions from \cite{Hurley2002}.

The single and binary systems thus generated are mixed in a proportion 74:25:1 for the thin and thick disk, and halo \citep{Torres2019}, astrometric and photometric magnitudes are derived in the {\it Gaia} passband filters and the whole population is normalized to the space density estimate of $4.9\times 10^{-3}\,{\rm pc^{-3}}$ as derived from the {\it Gaia} 100\,pc sample by \cite{Jimenez-Esteban18}. Finally, in order to mimic the observational results, we introduced a photometric and an astrometric error for each of the objects of our simulated sample based on {\it Gaia}'s performance\footnote{http://www.cosmos.esa.int/web/gaia/science-performance}.

\section{The observed {\it Gaia}-EDR3 100 pc white dwarf population}
\label{s:obs}

The high-accuracy astrometry and photometry provided by $Gaia$ allows us to separate different kind of sources in the Hertzsprung--Russell (HR) diagram, specially those located at short distances and with small parallax errors. Additionally, it is possible to identify reliable comoving sources by searching for sources with similar kinematics.

\subsection{Search methodology}

Recently, \cite{Jimenez-Esteban18} demonstrated that the white dwarf population accessible by \G\ up to 100 pc is nearly complete for a relative parallax error of 10\%. Thus, we search for objects within 100 pc in the \G-EDR3  catalogue\footnote{http://gea.esac.esa.int/archive/} by identifying all sources with accurate astrometric solutions and photometry. Then, the following criteria were applied:
\begin{itemize}
\item   $\varpi- 3*\sigma_{\varpi}\ge10$ 
\item $F_{\rm BP}/\sigma_{F_{\rm BP}}\ge10$ and $F_{\rm RP}/\sigma_{F_{\rm RP}}\ge10$
\item $|\mu_{\alpha}/\sigma_{\mu_{\alpha}}|\ge 10$ and $|\mu_{\delta}/\sigma_{\mu_{\delta}}|\ge 10$
\item ${\it RUWE} < 1.4$ 
\end{itemize}

\noindent where $\varpi$ is the parallax in mas; $F_{\rm BP}$ and $F_{\rm RP}$ are the fluxes and $G_{\rm BP}$ and $G_{\rm RP}$ the magnitudes in the bandpass filters BP and RP, respectively; $(\mu_{\alpha},\mu_{\delta})$ are the components of the proper motions in equatorial coordinates; {\it RUWE} (re-normalised unit weight error)\footnote{https://www.cosmos.esa.int/web/gaia/dr2-known-issues} is a parameter which basically tells us if the astrometric solution is reliable; and finally the $\sigma$ values are the errors of the corresponding parameters. 

\begin{figure}
\includegraphics[width=\columnwidth]{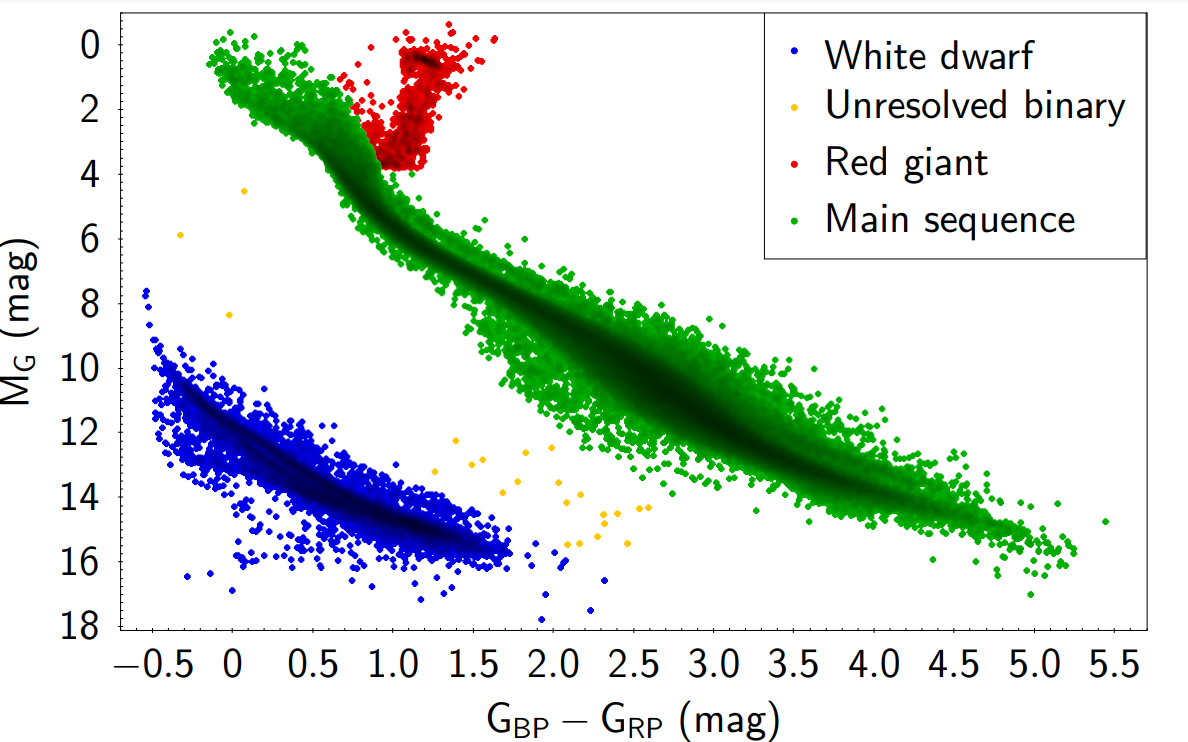}
\caption{Hertzsprung--Russell diagram of all Gaia sources within 100 pc considered in this work.}
\label{f:HRD}
\end{figure}

The selected sample obtained this way contains more than 212,000 sources. We considered only sources not flagged as {\it duplicated\_source} in the catalogue, since this indicates probable astrometric or photometric problems. In addition, we followed the recommendations by \cite{Riello20} and we imposed a threshold of 3$\sigma$ in the {\it corrected BP/RP excess factor}. Thus, the sample was reduced to 197,502 {\it Gaia} sources within 100 pc. Figure\,\ref{f:HRD} shows these sources in the HR diagram. We roughly split the sample in four groups based on the position of the source in this HR-diagram: main-sequence stars (184,318), red giants (892), white dwarfs (12,263) and unresolved binaries (23). Note that this classification did not pretend to be exhaustive and rigorous, but just a rough classification. Note also that the sources classified as unresolved binaries are very likely to belong to the WDMS or cataclysmic variable types since these objects are located between the white dwarf and main sequence loci \citep{Rebassa-Mansergas2021}.

The sample of 12,263 sources classified as white dwarfs represents {$\sim60-70\%$} of the total population of white dwarfs within 100\,pc, while the expected contamination is estimated to be lower than 1\% and mainly due to subdwarfs and cataclysmic variables \citep{Jimenez-Esteban18}.

\begin{figure}
  \includegraphics[width=\columnwidth]{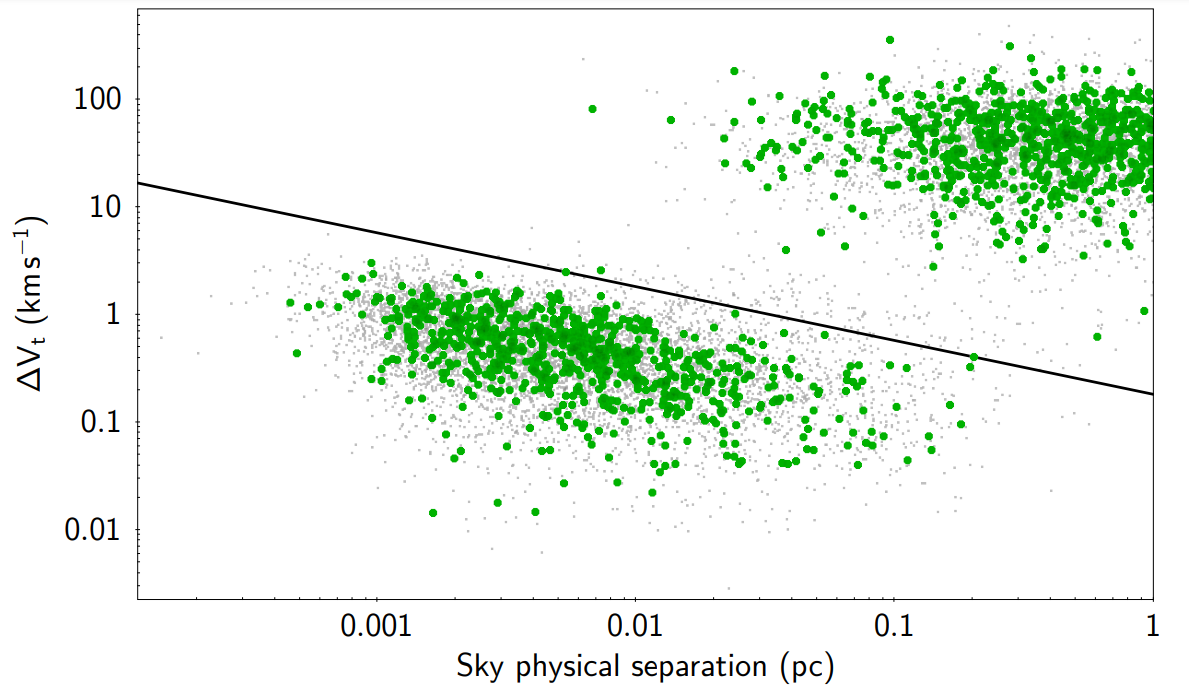}
   \includegraphics[width=\columnwidth]{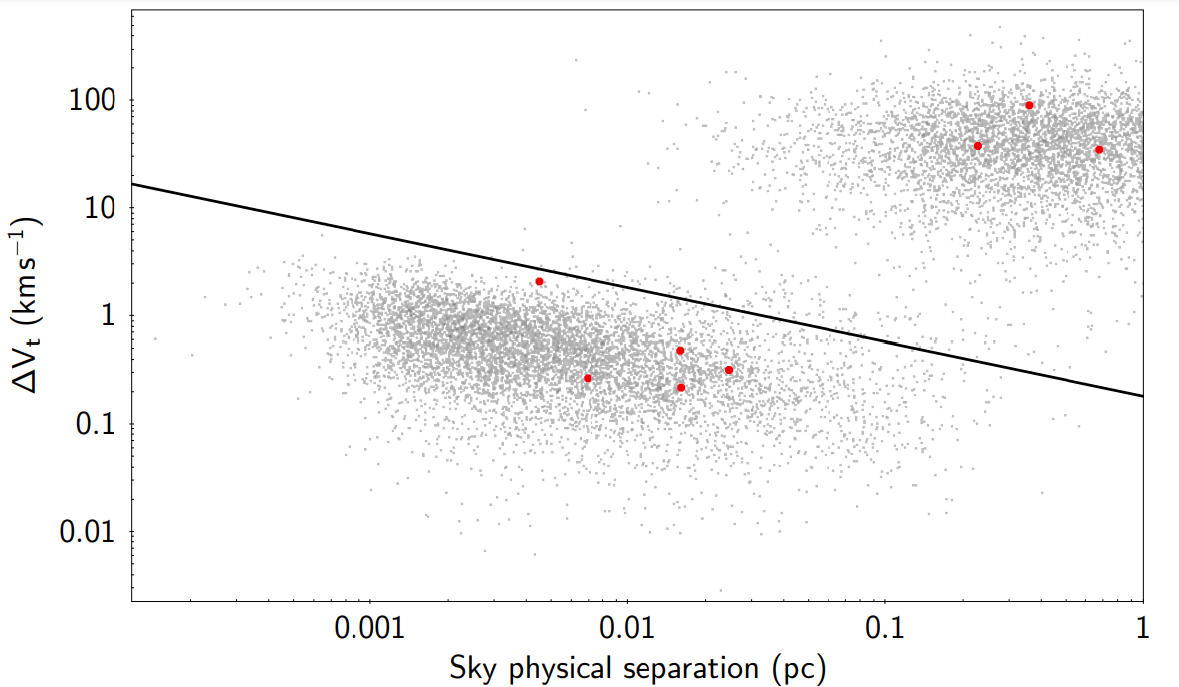}
     \includegraphics[width=\columnwidth]{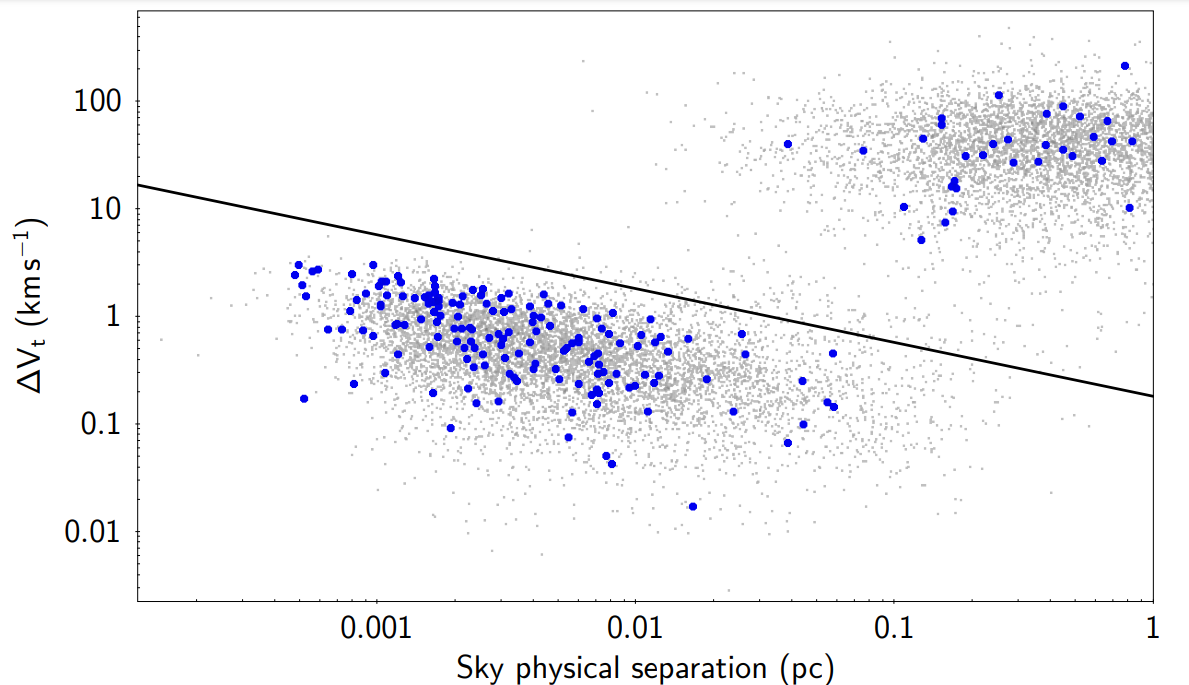}
\caption{Distribution of the difference in tangential velocity for all pair candidates (gray dots) as a function of the physical separation in the plane of the sky. Those systems which contain at least one white dwarf component are colored in green (white dwarf plus main sequence), red (white dwarf plus red giant) and blue (double white dwarf systems) and showed in the top, middle, and bottom panel respectively. The oblique continuous line represents the maximum difference in tangential velocity for a binary of $1.4+2.5$ \Msun\ stars on a circular orbit.
See text for details.}
\label{f-sel}
\end{figure}

\subsection{Identification of comoving pairs}
\label{ss:como}

The excellent astrometry of \G-EDR3 permits to search for binary systems containing at least one white dwarf. In the past, searches for co-moving stars were mainly based in the comparison of the proper motions. Now, \G-EDR3 provides not only accurate proper motions but precise parallaxes as well. Thus, we based our search of white dwarf binary systems in their separation in the space and their motion.

The search for co-moving star systems is usually affected by contamination by chance alignments (stars that show the same motion within the errors but that are actually not physically related). The effect increases with the physical separation of the candidate system \citep{Jimenez-Esteban2019}. Binary stars with larger separations than 1 pc have been reported in the literature \citep{Oh17,Oelkers17}. However, binaries with so large separations are not expected to survive long since they are more easily disrupted by close encounters with other stars or clouds of gas, as they move through the Galaxy. So, the probability to find a binary system with separation larger than 1 pc is very low, and at the same time, the expected contamination by chance alignments is very high \citep{Jimenez-Esteban2019}. Consequently, we imposed a maximum separation of 1 pc in both the plane of the sky ($D_t$\,<\,1 pc) and in the radial direction taking into account the error ($D_r-3\cdot\sigma_{D_r}$\,<\,1 pc) .

Since we searched for binaries within only 100\,pc from the Sun, instead of comparing the proper motion of the studied objects, we selected our binary systems by comparing their tangential velocities. Thus, for each of the selected sources within 100 pc and good astrometry, we calculated the difference in the tangential velocities with any other star located closer than 1 pc within the errors. Figure\,\ref{f-sel} shows the distribution of the differences in the tangential velocity for all the candidate pairs (gray dots) found as a function of the physical separation in the plane of the sky. We colored those systems with at least one white dwarf component. WDMS are marked in green in the top panel of Fig.\,\ref{f-sel}, if the companion is a red giant (WDRG) they are then plotted in red in the middle panel, while DWDs are shown as blue dots in the bottom panel. 

The distribution of pairs in Fig.\,\ref{f-sel} clearly follows two clusters: those pairs expected to be real, with small sky physical separation and low tangential velocity differences (bottom-left cluster), and those casual alignments, with high separation and high tangential velocity differences (top-right cluster). 
We also plotted an oblique continuous line representing the maximum difference in tangential velocity for a binary system of $1.4+2.5\,$\Msun stars (the maximum mass expected for a white dwarf and for a main-sequence A star within 100 pc, respectively) on a circular orbit. From Fig.\ref{f-sel} we identified as very likely real wide binary systems those located under that line, that is in the first cluster (bottom-left). This way we selected 768 WDMS co-moving pairs, 5 WDRG pairs, and 155 double degenerate systems.

\begin{figure}
\includegraphics[width=1.2\columnwidth,trim=47 60 0 60]{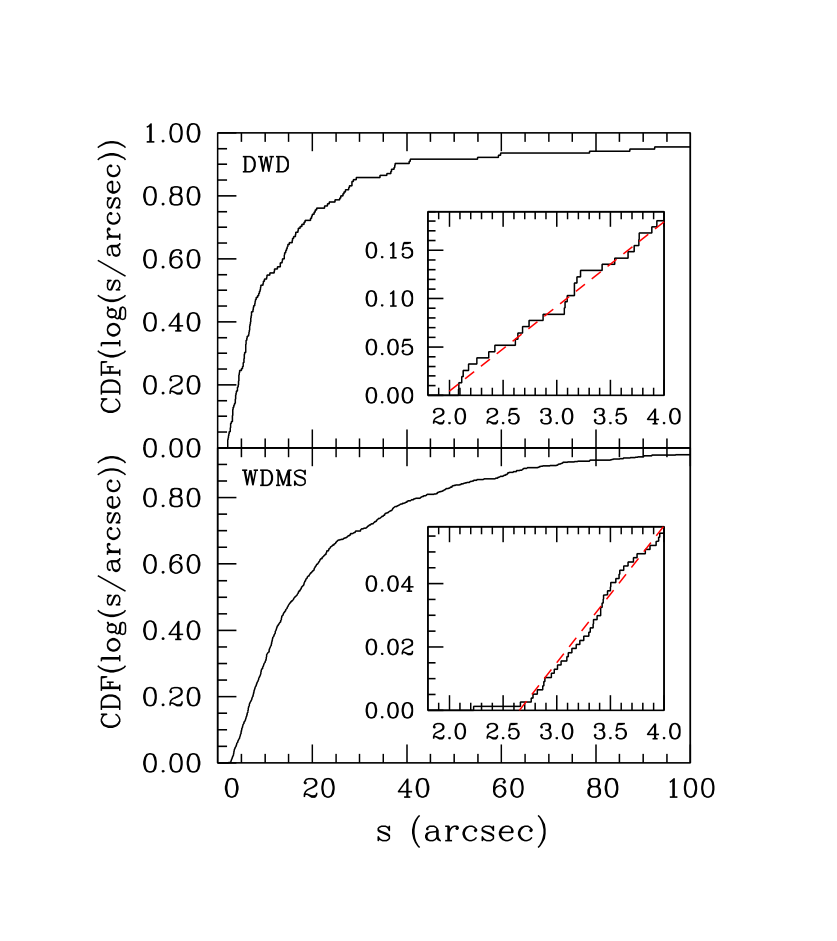}
\caption{Cumulative distribution function of the angular separations of our selected sample of WDMS (bottom panel) and DWD (top panel). A linear fitting (dashed red line in the zoomed in panels) reveals a minimum angular separation of $1.95\arcsec$ for DWD and $2.63\arcsec$ for WDMS.}
\label{f:sep}
\end{figure}

To be able to later compare with our synthetic samples we need to know the minimum angular separation above which a system can be considered as resolved. Nominally, {\it Gaia} performance setups this value at $\sim0.5\,\arcsec$ \citep{GaiaDR12016}. However, this value can be larger for a particular sample (as it is our case with astrometric and photometric selection criteria). In order to estimate this value, we plot in Fig.\,\ref{f:sep} the angular separation cumulative distribution function for our sample of WDMS (bottom panel) and DWD (top panel). A linear fitting (dashed red line in the zooming panels) reveals a minimum angular separation of $1.95\arcsec$ for DWD and $2.63\arcsec$ for WDMS. Although, in this last case, we have found some WDMS systems with lower angular separations, we can not ensure completeness below that value. Consequently, we adopt a conservative value of $2.75\arcsec$ as our boundary limit for both, WDMS and DWD, resolved systems. It is worth saying that this relatively large minimum angular separation value will provide a lower bound of the fraction of resolved systems but, at the same time, allow us to avoid possible effects of contrast ratio among binary components \citep[see][]{Toonen2017} specially important for very low angular separations.

Thus the final number of co-moving pairs understood as resolved WDMS is slightly reduced to 765, which represents $6.31\pm0.23\%$ of the total observed (single plus binary) white dwarf population, while the final number of resolved double degenerates results to be 143, accounting for $1.18\pm 0.10\%$, where errors have been estimated assuming a Poissonian distribution. The resolved white dwarf plus red giant stars sample is only marginal and represents $<\,0.1$\%. 

In a detailed analysis, \citet{ElBadry2018} by means of the photometry, proper motions and parallaxes derived from {\it Gaia}-DR2 within 200 pc from the Sun, found 177 pairs of double degenerate systems and 977 WDMS binaries within 100\,pc. Although these numbers of systems are slightly larger than the ones found here due to our different selection criteria, we should consider the percentages, that is the number of systems with respect to the total number of objects analyzed. The analysis done by \cite{ElBadry2018} does not provide this last value, however, a rough estimate, taking into account the general selection criteria applied in their work, shows that in their process the completeness of the total white dwarf population within 100\,pc is $\approx$75\%. Hence, the percentages (number of objects) of resolved WDMS and DWD we deduced from \citet{ElBadry2018} are 7.11\% (980) and 1.29\% (176), respectively, which are in very good agreement with our estimates. Finally, the percentage of resolved DWD found for the 100\,pc sample is, although slightly smaller, in agreement with the 1.7\% found in the 20\,pc sample by \cite{Toonen2017}. However, that is not the case for the resolved WDMS systems, where the value found for the 100\,pc sample is nearly half of the value estimated in the 20\,pc sample ($\approx$15\%). Probably some selection effects may account for this, as the physical separation depends on the distance and, consequently, a greater percentage of resolved systems is expected for closer samples. A more detailed analysis of this issue is presented in Section \ref{s:dis}.


\section{Results}
\label{s:res}
As previously indicated, our aim is to find a set of parameters for the binary synthetic population such that the resulting percentages of the resolved DWD and WDMS are compatible with observed values. In order to achieve this we proceed in two steps. First, we make a coarse analysis of the space parameters by identifying those that are relevant in the fitting process in the sense that the synthetic percentages achieve closer values to the observed ones. Then a fine-tuning process is performed that leads us to obtain a best-fit model.

\subsection{Preliminary analysis of the space parameter}
\label{s:mod}

\begin{table*}
  \begin{tabular}{| c | c | c | c | c | c | c | c | c | c |c | c |}
  \hline
  Model & $f_b$ & $n(q)$ & $f(a)$ & WD$_{\rm sin}$ & WD$_{\rm mer}$ & DWD$_{\rm res}$ & WDMS$_{\rm res}$ & DWD$_{\rm unres}$ & WDMS$_{\rm unres}$ & WDRG$_{\rm res}$ & $D\pm\sigma_{\rm D}$ \\ \hline
  1 & 0.50 & 1 & $a^{-1}$ & 56.82 & 25.52 & 5.19 & 6.66 & 5.24 & 0.32 & 0.25 & $4.03\pm0.11$ \\ 
  2 & 0.50 & $q$ & $a^{-1}$ & 54.90 & 26.19 & 7.43 & 4.45 & 6.62 & 0.05 & 0.37 & $6.52\pm0.15$  \\ 
  3 & 0.50 & $q^{-1}$ & $a^{-1}$ & 58.02 & 25.94 & 2.27 & 10.30 & 2.02 & 1.31 & 0.14 & $4.14\pm0.22$  \\ \hline
   4 & 0.35 & 1 & $a^{-1}$ & 71.01 & 17.13 & 3.49 & 4.47 & 3.51 & 0.21 & 0.18 & $2.96\pm0.13$ \\ 
  5 & 0.35 & $q$ & $a^{-1}$ & 69.37 & 17.82 & 5.03 & 3.02 & 4.48 & 0.04 & 0.25 & $5.06\pm0.13$ \\ 
  6 & 0.35 & $q^{-1}$ & $a^{-1}$ & 71.86 & 17.43 & 1.51 & 6.84 & 1.36 & 0.89 & 0.10 & $0.64\pm0.17$  \\ \hline
   7 & 0.50 & 1 & BPL  & 54.38 & 38.41 & 1.46 & 1.75 & 3.64 & 0.30 & 0.05 & $4.56\pm0.10$  \\ 
  8 & 0.50 & $q$ & BPL  & 53.18 & 39.30 & 1.78 & 0.91 & 4.59 & 0.17 & 0.07 & $5.43\pm0.06$ \\ 
  9 & 0.50 & $q^{-1}$ & BPL & 56.58 & 37.55 & 0.57 & 2.75 & 1.38 & 1.13 & 0.03 & $3.61\pm0.11$  \\ \hline 
     10 & 0.85 & 1 & BPL  & 17.38 & 69.47 & 2.68 & 3.20 & 6.63 & 0.55 & 0.09 & $3.45\pm0.05$  \\ 
  11 & 0.85 & $q$ & BPL & 17.08 & 71.34 & 3.22 & 1.64 & 8.31 & 0.32 & 0.12 & $5.20\pm0.11$  \\ 
  12 & 0.85 & $q^{-1}$ & BPL & 19.07 & 71.71 & 1.11 & 5.27 & 2.68 & 2.13 & 0.06 & $1.17\pm0.09$ \\\hline 
      13 & 0.50 & MDS & $a^{-1}$ & 55.82 & 27.72 & 5.20 & 6.06 & 4.55 & 0.32 & 0.33 & $4.03\pm0.14$ \\
    14 & 0.50 & MDS & BPL & 55.02 & 38.88 & 1.38 & 1.46 & 2.84 & 0.28 & 0.13 & $4.85\pm0.08$ \\ 
   \hline 
         15 & 0.35 & MDS & $a^{-1}$ & 70.10 & 18.76 & 3.54 & 4.10 & 3.07 & 0.22 & 0.22 & $3.24\pm0.15$ \\
    16 & 0.85 & MDS & BPL & 18.14 & 72.59 & 2.57 & 2.72 & 5.27 & 0.51 & 0.25 & $3.97\pm0.10$ \\ 
   \hline 

  \end{tabular}

\caption{Preliminary models evaluated to determine a first approach to the best fit model, i.e., the model that simulates DWD$_{\rm res}$ (\%) and WDMS$_{\rm res}$ (\%) closest to the observed values. For all models here we assume $\alpha_{\rm IMF}=-2.3$.}
\label{t:models_fns}
\end{table*}

Among the numerous parameters that shape the binary sample of our population synthesis code, we left as free parameters those which are relevant for the final number of resolved systems. In this sense, we initially disregard the initial eccentricity distribution and the common envelope treatment. We recall that in the first case we used a standard thermal distribution \citep{Heggie1975} and for the second one we adopted two models based on the $\alpha$-formalism prescription and one model based in the $\gamma$ prescription (see Section \ref{s:ce}). In any case, the physical separation of resolved binary systems is large enough so that their fraction respect to the total populations are not affected by these assumptions. In particular, the different treatment adopted of the common envelope phase changes the percentage of unresolved systems (see Section \ref{s:ce}), and hence white dwarfs as product of mergers, but not the number of resolved binaries. Consequently, we first explore a variety of models changing only a set of selected parameters (those that are expected to have a larger impact in the percentages of resolved WDMS and DWD binaries), with the aim of narrowing our selection down to only two parameters. The four initial free parameters we select are:
\begin{enumerate}
  \item The initial binary fraction: $f_b$. This parameter gives us the probability that an object is originally generated as a single star or it belongs to a binary system. Although a value of $f_b=0.50$ is generally adopted, a wider range of values ($0.35 \le f_b \le 0.85$) based on star counts of different stellar types can be found in the literature \citep[see][and references therein]{DucheneKraus2013,Moe2017,Toonen2017,Niu2021}.
  \item The initial mass ratio distribution (IMRD): The IMRD, $n(q)$, has been explored in previous studies with the population synthesis code presented here \citep[e.g.][]{Cojocaru2017,Torres2018}. These studies, although not conclusive, point out that an inverse relation between the mass of the secondary star in the binary system with respect to the primary star has a better performance that other approaches at reproducing the observational data for the binary white dwarf population. However, this could not be the case for other kind of binary populations \citep[see, for instance,][]{Moe2017}. We revisit here this issue by evaluating different models of the $n(q)$ function.
  \item The initial separation distribution, $f(a)$, of main-sequence binaries. Recent studies \citep[e.g.][]{ElBadry2018} show that the final separation distribution for DWDs, WDMS, or main-sequence binary systems present different behaviours. This fact reveals that a common initial separation distribution suffers from the binary evolution and, consequently, different separation models shall be considered.
  \item The slope of the initial mass function (IMF): $\alpha_{\rm IMF}$. Although the IMF can be initially considered as a universal function and its slope reasonably constrained, several studies reveal that this function can suffer from local inhomogeneities (as shall be the case for our sample of 100 pc) or specific slopes for different stellar types \citep[e.g][]{Bastian2010,Sollima2019}.
\end{enumerate}

The results corresponding to our different models are displayed in 
Tables \ref{t:models_fns}, \ref{t:models_ans}, and \ref{t:models_ce}.
The first four columns refer to the model number and three of the four parameters mentioned above, respectively. The rest of the columns are indicated as follows: 
\begin{description}
  \item WD$_{\rm sin}$: the fraction of white dwarfs that come from single stellar evolution.
  \item WD$_{\rm mer}$: the fraction of white dwarfs resulting from the merger of both stars in a binary system.
  \item DWD$_{\rm res}$: the fraction of resolved double white dwarf binaries.
  \item WDMS$_{\rm res}$: the fraction of resolved WDMS binaries.
  \item WDRG$_{\rm res}$: the fraction of resolved WDRG binaries.
  \item DWD$_{\rm unres}$: the fraction of unresolved double white dwarf
  \item WDMS$_{\rm unres}$: the fraction of unresolved WDMS binaries. It is worth saying here that we refer to such an object if it lies on the white dwarf single region of the HR-diagram. That implies that the bolometric magnitude of the white dwarf should be brighter than its main-sequence counterpart. Otherwise, the binary object will be located outside this region, mainly on the main-sequence path of the HR-diagram \citep[see][]{Rebassa-Mansergas2021} and, thus, not considered in our sample.

\end{description}

The sum of columns 5 to 11, i.e. WD$_{\rm sin}$+WD$_{\rm mer}$+DWD$_{\rm res}$+ WDMS$_{\rm res}$+DWD$_{\rm unres}$+WDMS$_{\rm unres}$+WDRG$_{\rm res}$ is the full 100\% population. It is worth noting here that these percentages of the different sub-populations are relative to the total population of white dwarfs within 100\,pc in the region of the HR-diagram where single white dwarfs are expected to be found \citep[see][]{Jimenez-Esteban18}. It is also important to remark that we are counting systems, that is DWD systems are counted as one.

Finally, and for a more quantitative comparison of models with observed data, we include in the last columns of the tables 
the value of the estimator $D$, which is defined as the euclidean distance to the observed percentages of resolved WDMS ($6.31\%$) and resolved DWD ($1.18\%$):

\begin{equation}
  D=\sqrt{(\rm{DWD_{\rm res}(\%)}-1.18)^2+(\rm{WDMS_{\rm res}(\%)}-6.31)^2}
  \label{eq:estim}
\end{equation}

Taking into account the errors in the observed percentages, we can propagate them in the estimator $D$. Thus  $1\sigma$, $2\sigma$ and $3\sigma$ errors in the observed parameters correspond to distance estimator values of 0.37, 0.74 and 1.11, respectively. In other words, the distance $D$ is inversely proportional to the confidence level of a given model.

We start with our reference model, Model 1, consisting in a flat IMRD distribution, $n(q)\propto 1$, an inverse initial orbital separation, $f(a)\propto a^{-1}$, and a binary fraction of $f_{b}=0.50$. We then considered two additional IMRDs and maintaining fixed the rest of the parameters. Hence, Model 2 and 3, refer then to the cases of a IMRD $n(q)\propto q$ and $n(q)\propto q^{-1}$, respectively. Although the number of IMRD proposed in the literature is much larger, \citep[see, for instance,][for a recompilation of up to 12 different IMRD models]{Cojocaru2017}, here we are now only interested in the general trend of the IMRD. We find out that the DWD resolved fractions arising from the simulated samples are relatively larger than the observed value of 1.18\%, being Model 3 with a $n(q)\propto q^{-1}$ distribution the one with a lowest value of $2.27\%$. We also observe that the ratio of $\rm{WDMS_{\rm res}}/\rm{DWD_{\rm res}}$ is close to 1 in Model 1, smaller than 1 in Model 2, and around 5 for Model 3, in agreement with the observed ratio only in the later case. However, all the models present a rather large $D$ estimator. In Models 4 to 6 we replicate Models 1 to 3, respectively, but now looking for the binary fraction that minimizes the distance estimator $D$. The model with a $q^{-1}$ IMRD (Model 6) yields the smallest $D$ value, as indicative that the percentages of resolved DWDs and WDMS are in better agreement with the observed ones. However, all three models (included Model 6) present a fraction of binaries, $f_{\rm b}=0.35$, which is the minimum value analyzed. This indicates that probably a lower binary fraction can even better match the observed values. This possibility has been explored in the fine-tuning analysis of Section \ref{s:grid}.

In a next step,  we investigate the effects of varying the initial separation $a$. One of the preferred models for solar-type stars is a log-normal distribution of periods \citep{Raghavan2010}. However, this model seems not to produce a smaller number of resolved double degenerate systems than its counterpart the inverse $a^{-1}$ distribution \citep{Toonen2017}. Instead we use a power-law distribution, $N(a)\propto a^{n_a}$ with $n_a=-1.3$, which seems to better fit the final DWD observed distribution at the low separation range $a<0.05$ au \citep{Maoz2017,Maoz2018}. Recently, \citet{ElBadry2018} proposed a variation of this model, named broken power-law (BPL) model, consisting of a power-law distribution with different exponents $n_a$ depending on the separation region. In any case, it is expected that a steeper slope of the semimajor axis distribution will produce fewer resolved systems. Thus, we adopt the BPL model for the main-sequence distribution of stars \citep{ElBadry2018}. Basically, this new model consist in a power-law distribution with $n_a=-1.6$, up to separations of $\sim6 300$\,AU. For larger separations the BPL distribution assumes an even lower exponent ($n_a=-1.8$), but our simulations show that the effect on the final percentage of the resolved sub-populations is negligible. The results are shown in Table \ref{t:models_fns} in Models 7 to 9, as replicating Models 1 to 3, respectively, but now adopting a BPL distribution. The percentage of resolved systems is notably reduced in the three models, while the relative ratio is maintained as in Models 1 to 3. Consequently, the model with an inverse IMRD provided again the best fit to the observed value. In Models 10 to 12, we repeat the exercise as previously done, leaving the binary fraction as a free parameter. On the contrary to the previous models, the best matches are now achieved for the larger binary fraction under study, i.e., $f_{\rm b}=0.85$. Moreover, Model 12, which assumes an inverse IMRD achieves again a relative good agreement with the observed values. However, a major drawback is present in these solutions. The large percentage of white dwarfs mergers generated with these models, accounting for nearly $\sim70\%$ of the entire population, seems to be hardly realistic \citep[see, for instance,][where they provide an estimate  of $10-30\%$]{Temmink2020}. Although the fraction of mergers is basically related to the common-envelope treatment (see Section \ref{s:ce}), the use of a BPL model seems to provide worse results than the initial $a^{-1}$ separation models.

To further analyze the effect of the IMRD, we follow the recent analysis from \cite{Moe2017} of the mass-ration of binary stars as a function of the period of the orbit. Basically, we assume a solar-type model, with a null $n(q)$ for ratios $q<0.1$ accounting for the brown desert, an increasing $n(q)\propto q^{0.3}$ in the range $0.1<q<0.3$, and a decreasing $n(q)\propto q^{-0.5}$ up to the twin-regime, where an excess of these systems is included. The results, 
when the IMRD of \cite{Moe2017} is used (MDS) and
corresponding to a binary fraction $f_{\rm b}=0.50$ and the $a^{-1}$ and BPL separation model are shown in Table \ref{t:models_fns} as Models 13 and 14, respectively. The results obtained for Models 13 and 14 are similar to those of Models 3 and 9, respectively. Consequently, regardless of the separation model, the MDS model seems to be analogous to the $q^{-1}$ model. We further explore this by leaving the binary fraction as a free parameter. The corresponding Models 13 and 14 are shown in Table \ref{t:models_fns} as Models 15 and 16, respectively. The fraction of resolved DWD and WDMS systems is slightly closer to the observed values, reproducing the same trend as previously observed, i.e., the $a^{-1}$ separation model achieves the best match for the minimum binary fraction, while the reverse happens for the BPL model.

The evaluation so far of the resulting proportions of resolved DWD and WDMS systems can be summarized in the following four main conclusions:
\begin{enumerate}
  \item The ratio of resolved WDMS to DWD systems mainly depends on the IMRD adopted, being the $n(q)\propto q^{-1}$
  the model that better replicates the observed data.
  \item A BPL for the initial orbital separation of the binary systems yields reasonable low values of resolved DWD systems in good agreement with the observed percentage. However, in order to achieve a resolved WDMS percentage similar to the observed one, it requires a large binary fraction which results in an unrealistic proportion of white dwarf mergers.
  \item Adopting a more sophisticated IMRD following \citet{Moe2017} provides worse fits than those obtained using the $n(q)\propto q^{-1}$ model, regardless of the orbital separation model.
  \item The best models found so far has been achieved for a low binary fraction.
\end{enumerate}

\subsection{The sensitivity of the initial-mass function slope} 
\label{s:imf}

\begin{table*}
  \begin{tabular}{| c | c | c | c | c | c | c | c | c | c | c |c | c |}
  \hline
  Model & $\alpha_{\rm IMF}$ & $n(q)$ & $f(a)$ & WD$_{\rm sin}$ & WD$_{\rm mer}$ & DWD$_{\rm res}$ & WDMS$_{\rm res}$ & DWD$_{\rm unres}$ & WDMS$_{\rm unres}$ & WDRG$_{\rm res}$ & $D\pm\sigma_{\rm D}$ \\ \hline
 
  17 & -2.2 & 1 & $a^{-1}$ & 55.27 & 26.69 & 5.85 & 6.36 & 5.28 & 0.24 & 0.31 & $4.67\pm0.15$ \\ 
   1 & -2.3 & 1 & $a^{-1}$ & 56.82 & 25.52 & 5.19 & 6.66 & 5.24 & 0.32 & 0.25 & $4.03\pm0.11$ \\ 
  18 & -2.4 & 1 & $a^{-1}$ & 58.44 & 25.30 & 5.11 & 5.93 & 4.59 & 0.36 & 0.27 & $3.95\pm0.13$ \\ \hline
 19 & -2.2 & $q^{-1}$ & $a^{-1}$ & 55.92 & 27.42 & 2.08 & 11.17 & 2.28 & 0.94 & 0.19 & $4.94\pm0.20$  \\ 
  3 & -2.3 & $q^{-1}$ & $a^{-1}$ & 58.02 & 25.94 & 2.27 & 10.30 & 2.02 & 1.31 & 0.14 & $4.14\pm0.22$ \\ 
  20 & -2.4 & $q^{-1}$ & $a^{-1}$ & 59.53 & 25.30 & 1.84 & 10.26 & 1.93 & 0.99 & 0.15 & $4.01\pm0.21$ \\ \hline 
  21 & -2.2 & MDS & $a^{-1}$ & 54.14 & 28.76 & 5.56 & 6.11 & 4.77 & 0.34 & 0.33 & $4.39\pm0.13$ \\ 
  13 & -2.3 & MDS & $a^{-1}$ & 55.82 & 27.72 & 5.20 & 6.06 & 4.55 & 0.32 & 0.33 & $4.03\pm0.14$ \\ 
  22 & -2.4 & MDS & $a^{-1}$ & 57.50 & 26.85 & 4.93 & 5.78 & 4.27 & 0.36 & 0.32 & $3.79\pm0.17$ \\ \hline
   
  \end{tabular}

\caption{Preliminary models evaluated to determine a first approach to the best fit model, i.e., the model that simulates DWD$_{\rm res}$ (\%) and WDMS$_{\rm res}$ (\%) closest to the observed values. For all models here we assume $f_{\rm b}=0.5$.}
\label{t:models_ans}
\end{table*}

Although the initial mass function is generally accepted as a universal distribution, the different models presented in the literature seem to claim that this is not the case \citep[see, for a recent and comprehensive analysis of this issue][]{Sollima2019}. The mass range for single white dwarf progenitors lies in the range $\sim 0.5$ up to $\sim 9\,$\Msun. However, when binary evolution is considered , this range can be extended. Nevertheless, we will maintain a constant slope for masses greater that $\sim 0.5\,$\Msun. In our reference model, as those analyzed in the previous section, we adopt a standard value of $\alpha_{\rm IMF}=-2.3$. Now, we check the sensitivity of the initial-mass function slope, by varying it from -2.2 to -2.4 for three of our previous models. We select Model 1 as our reference model, and Models 3 and 13 as the two models that, for a fixed $f_{\rm b}=0.50$, present the best fittings achieved so far. In other words, we disregard the $n(q) \propto q$ model and the BPL model for the orbital separation. The former because it produces an inverse ratio of resolved WDMS with respect to DWD, and the latter because it generates an excessive number of white dwarf mergers. The corresponding new models are presented in Table \ref{t:models_ans}. We can observed that the change in the initial-mass function slope only produce a slight change in the results. In all cases a steeper slope yields a better agreement, although it is not enough to reproduce the observed values. Moreover, the change in the initial-mass function slope does not seem to vary the ratio of WDMS to DWD systems. As a conclusion we can state that a moderate change in the initial-mass function in the range of solar-type stars only produced a slight variation in the relative fraction of white dwarf populations. Therefore we can confirm that fixing $\alpha_{\rm IMF}$ to -2.3 is a safe assumption.

\subsection{The effect of the common-envelope treatment}
\label{s:ce}

\begin{table*}
  \begin{tabular}{| c | c | c | c | c | c | c | c | c | c | c |c | c |}
  \hline
  Model & CE-prescription & $n(q)$ & $f(a)$ & WD$_{\rm sin}$ & WD$_{\rm mer}$ & DWD$_{\rm res}$ & WDMS$_{\rm res}$ & DWD$_{\rm unres}$ & WDMS$_{\rm unres}$ & WDRG$_{\rm res}$ & $D\pm\sigma_{\rm D}$ \\ \hline
 
  1 & $\alpha_{\rm CE}=0.3$ & 1 & $a^{-1}$ & 56.82 & 25.52 & 5.19 & 6.66 & 5.24 & 0.32 & 0.25 & $4.03\pm0.11$ \\ 
  23 & $\alpha_{\rm CE}=1$ & 1 & $a^{-1}$ & 58.47 & 23.44 & 5.77 & 6.38 & 5.27 & 0.35 & 0.31 & $4.60\pm0.14$ \\ 
    24 & $\gamma=1.75\,\alpha\lambda=2$ & 1 & $a^{-1}$ & 60.63 & 14.83 & 6.00 & 6.63 & 11.04 & 0.52 & 0.35 & $4.83\pm0.14$ \\ \hline
  3 & $\alpha_{\rm CE}=0.3$ & $q^{-1}$ & $a^{-1}$ & 58.02 & 25.94 & 2.27 & 10.30 & 2.02 & 1.31 & 0.14 & $4.14\pm0.22$ \\ 
 25 & $\alpha_{\rm CE}=1$ & $q^{-1}$ & $a^{-1}$ & 59.80 & 23.76 & 2.08 & 11.07 & 2.26 & 0.88 & 0.16 & $4.85\pm0.21$ \\ 
  26 & $\gamma=1.75\,\alpha\lambda=2$ & $q^{-1}$ & $a^{-1}$ & 63.83 & 14.75 & 2.22 & 11.81 & 5.76 & 1.47 & 0.16 & $5.59\pm0.24$ \\ \hline 
    13 & $\alpha_{\rm CE}=0.3$ & MDS & $a^{-1}$ & 55.82 & 27.72 & 5.20 & 6.06 & 4.55 & 0.32 & 0.33 & $4.03\pm0.14$ \\ 
  27 & $\alpha_{\rm CE}=1$ & MDS & $a^{-1}$ & 57.43 & 25.27 & 5.35 & 6.24 & 5.04 & 0.33 & 0.34 & $4.18\pm0.16$ \\ 
  28 & $\gamma=1.75\,\alpha\lambda=2$ & MDS & $a^{-1}$ & 58.44 & 16.42 & 5.45 & 6.35 & 12.68 & 0.33 & 0.34 & $4.27\pm0.17$ \\ \hline
   
  \end{tabular}

\caption{Preliminary models evaluated to determine a first approach to the best fit model, i.e., the model that simulates DWD$_{\rm res}$ (\%) and WDMS$_{\rm res}$ (\%) closest to the observed values. For all models here we assume $f_{\rm b}=0.5$ and $\alpha_{\rm IMF}=-2.3$.}
\label{t:models_ce}
\end{table*}

The common-envelope phase plays a capital role in the binary evolution, although the precise mechanisms that take place in it are not well understood. Several prescriptions have been proposed for their modelling. Even within the same prescription, the values of the different involved parameters are uncertain. However, we are not concerned about the precise description of the common-envelope phase, but rather on the effect on the relative populations arising from the simulations.

In order to check this, we analyze three different models for the common-envelope phase. The first two follow the so-called $\alpha_{\rm CE}$ treatment, for which we adopt the values $\alpha_{\rm CE}=0.3$ and $\alpha_{\rm CE}=1.0$. For our third model we use the $\gamma$ prescription and adopting the value of $\gamma=1.75$ and $\alpha\lambda=2$ as in \cite{Toonen2017}. We apply these three assumptions to our reference model (Model 1), and to the so far two best choices (Model 3 and Model 13) when fixing the binary fraction to 0.5 and $\alpha_{\rm IMF}$ to -2.3. The results are summarized in Table \ref{t:models_ce}.

An overall glance at the results shows that the percentages for the different resolved populations and, consequently, the estimators $D$, are quite insensitive to the common envelope prescription used. That is initially expected, given that resolved systems are wider enough so not to be affected by the particular modelling of close systems \citep[see, for instance][]{Toonen2017}. 

On the other hand, within the $\alpha-$formalism, a larger value of $\alpha_{\rm CE}$ induces a lower fraction of the orbital energy released due to the shrinkage of the orbit used in expelling the envelope. Consequently, those models with $\alpha_{\rm CE}=1$ (Models 23, 25 and 27) achieve larger orbits and thus, less merger episodes are produced. This effect is even more intense when the $\gamma-\alpha$ prescription is adopted. In this case, the first common-envelope episode is carried out through the $\gamma-$formalism, yielding an stable but not conservative phase. In case of a possible unstable second common episode, the $\alpha-$formalism is used. Because of this, the production of mergers in this formalism is rather limited, since a widening effect of the orbits has occurred during the first common envelope episode. That is reflected by the lowest percentage of mergers for Models 24, 26 and 28 shown in Table \ref{t:models_ce}.

\subsection{Fine-tuning parameter analysis: the best fit models.}
\label{s:grid}

\begin{table*}
  \begin{tabular}{| c | c | c | c | c | c | c | c | c | c | c |c | c |}
  \hline
  Model & $f_{\rm b}$ & $n_q$ & $n_a$ & WD$_{\rm sin}$ & WD$_{\rm mer}$ & DWD$_{\rm res}$ & WDMS$_{\rm res}$ & DWD$_{\rm unres}$ & WDMS$_{\rm unres}$ & WDRG$_{\rm res}$ & $D\pm\sigma_{\rm D}$ \\ \hline

29 & 0.32 & $-1.13$ & $-1$ & 79.12\,(79.47) &	16.36\,(8.76) &	1.19 &	6.66 &	1.22\,(2.94) &	0.65\,(0.92) &	0.06 & $0.12\pm0.06$\\
30& 0.30 & MDS & $-0.94$ & 79.25\,(79.97) &	13.30\,(4.61) &	3.85 &	5.31 &	3.13\,(5.82) &	0.19\,(0.21) & 0.23 & $2.73\pm0.12$ \\
\hline
  \end{tabular}
  
\caption{Set of parameters corresponding to the binary fraction, $f_{\rm b}$, the slope of the IMRD function, $n_q$ and the slope of the initial separation distribution, $n_a$, second to fourth columns, respectively, for our best models (Models 29 and 30). The rest of the columns indicate the percentages of the different sub-populations (see Section\,\ref{s:mod}) and those in brackets when a $\gamma$ prescription for the CE treatment is used (see Section \ref{s:ce}). Our Model 29 perfectly matches the observed values for the resolved populations of WDMS and DWD.}
\label{t:models_best}
\end{table*}

\begin{figure*}
  \includegraphics[width=0.95\columnwidth,trim=0 0 0 0, clip]{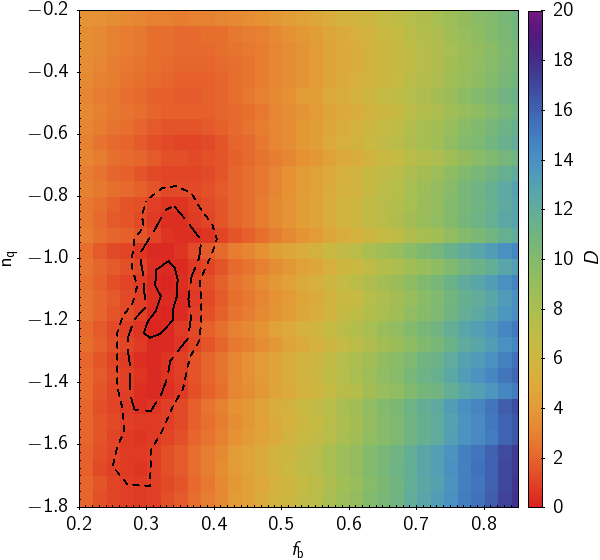}
  \hspace{0.5cm}
    \includegraphics[width=0.95\columnwidth,trim=0 0 0 0, clip]{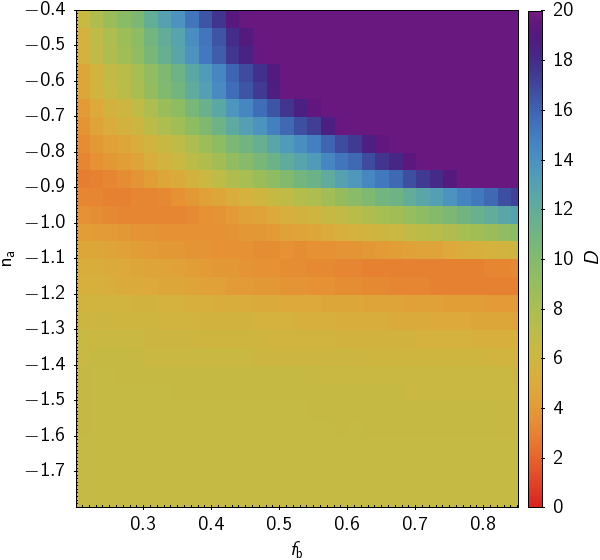}
     \caption{Density map of the distance estimator $D$ (color scale) as a function of the binary fraction, $f_{\rm b}$, and the IMRD slope, $n_q$, for Model 29 (left panel) and as a function of the binary fraction, $f_{\rm b}$, and the initial separation slope, $n_a$ for Model 30 (right panel). Also shown are the confident level contours for $1\sigma$, $2\sigma$ and $3\sigma$ marked by a continuous, dashed and dotted black line, respectively. An optimal solution is found for Model 29 with $f_{\rm b}=0.32$ and $n_q=-1.13$.}
  \label{f:dens}
\end{figure*}

From the analysis done so far we can conclude that the model with an inverse IMRD and a low binary fraction, that is Model 6, presents the best fit. We can disregard an increasing IMRD or an BPL initial separation model, as well as we can ensure that the effects of the IMF slope or the common envelope treatment are of minor order with respect to the percentage of resolved systems. Consequently, starting from Model 6, we explore the space parameter leaving the fraction of binaries, $f_b$, and the IMRD slope, $n_q$, as free parameters (Model 29). Although  models with an inverse IMRD slope are the one that presents best results so far, we also consider the alternative model (Model 30) with the IMRD from MDS, also leaving in this case the fraction of binaries, $f_b$, and the slope, $n_a$, of the initial separation distribution, as free parameters.

The corresponding results are shown in Fig.\,\ref{f:dens}, where we plot the density map of the space parameters as a function of the distance estimator $D$ depicted in a color scale. In the left panel we show the density map corresponding to Model 29. Regions within $1\sigma$, $2\sigma$ and $3\sigma$ levels are marked by a continuous, dashed and dotted black line, respectively. We recall that these confidence-level contours correspond to our distance estimator $D$ values of 0.37, 0.74 and 1.11, respectively. Hence, the region inside a certain contour line indicates the set of parameters of the model  that generates a result that is up to a certain distance from the observed values. The best fits are achieved for a slope of the IMRD function between $n_q=-1.0$ and $n_q=-1.2$ and a binary fraction of roughly $35\%$. We can also notice that a wider range of values for the slope from $n_q=-0.8$ up to $n_q=-1.7$ and a binary fraction in the range $f_b=0.25-0.40$ are compatible with the observed percentages within a $3\sigma$ confidence level. The corresponding density map for Model 30 is shown in the right panel of Fig. \ref{f:dens}. In this case we observe that there is no solution within the $3\sigma$ level. The minimum values of the distance estimator $D$ are found for a slope of the initial distribution $n_a\approx -1$ and a binary fraction of $\approx35\%$ and also for a steeper slope $n_a\approx -1.1$ and a binary fraction in the range $\sim 0.6-0.8$. However, in this last case, a large fraction ($\sim60\%$) of mergers is generated, as it happened when the BPL model was used, thus disregarding this possibility as a realistic solution. 

In Table \ref{t:models_best} we summarize the percentages for the different sub-populations derived from the best configuration of parameters for Models 29 and 30. We check that for Model 29 we can find a set of parameters that perfectly matches the observed values for resolved systems. The binarity fraction of our best-fit, although lower than previous estimates, is in agreement at the $2\sigma$ confidence level (27\%-39\%) with the recent estimates from {\it Gaia}-EDR3 and LAMOST data release 5 \citep{Niu2021}, where they claim a 39\% for late G and early K thick disk dwarf stars. That is not the case for Model 30, where the optimum solution, although it yields a reasonably low $D$ value, is far from being as good as Model 29.

\subsection{By-products of the best fit models}

The set of input parameters used in building the synthetic population for the best fit Models 29 and 30 (although the best fit for Model 29 perfectly matches the observed percentages, we continue referring to the best fit Model 30 for comparative purposes) are in agreement with the observed percentages of resolved populations. In this sense, our initial goal has been accomplished. However, we extend our analysis to the distributions of other parameters of the resulting synthetic populations. A detailed analysis will be beyond the objective of the present work, in which we focus in four important cases: the He-core white dwarf fraction, the final orbital separation distribution, the mass distribution and the fraction of resolved systems as a function of the sample volume.

\subsubsection{The fraction of Helium-core white dwarfs}

\begin{figure}
\includegraphics[width=1.1\columnwidth,trim=12 50 0 90, clip]{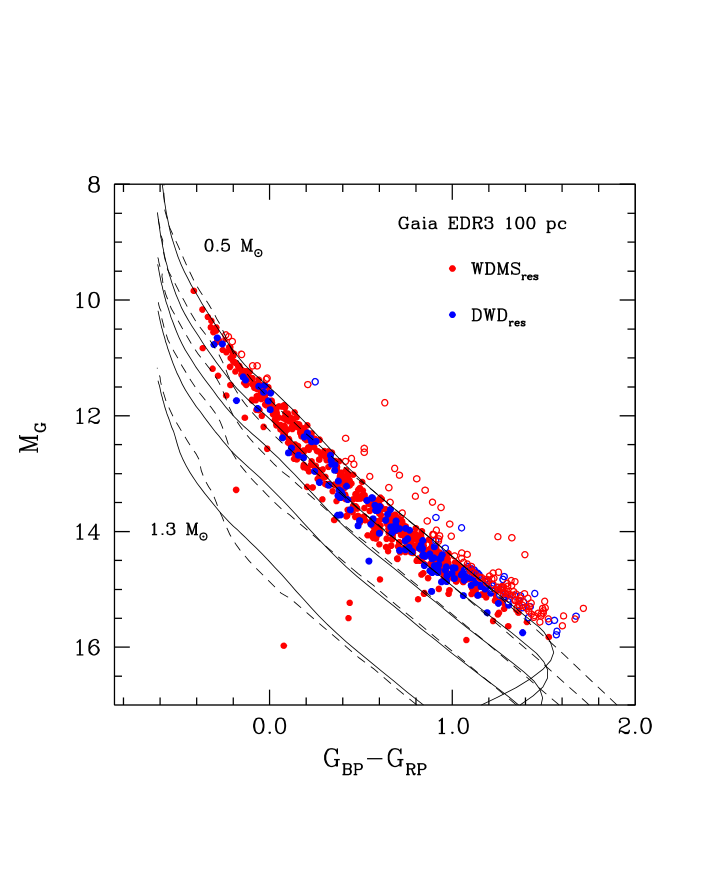}
\caption{Hertzsprung-Russel diagram of resolved WDMS (red dots) and DWD (blue dots) systems considered in this work. Marked as open symbols those with masses below $0.5\,$\Msun, hence helium-core white dwarf candidates. For illustrative purposes we also show the  cooling tracks for different masses in steps of 0.2\,\Msun for hydrogen-pure (black solid line) and hydrogen-deficient (black dashed line) white dwarf models from \citet{Camisassa2019} and references therein.}
\label{f:hewd}
\end{figure}

One of the by-products of our populations synthesis analysis is the percentage of He-core white dwarfs generated by our models. In Table \ref{t:He-WD} we show for our two best fit models, the percentages of He-core white dwarfs with respect to the corresponding sub-population, that is resulting from resolved DWD systems (second column), resolved WDMS systems (third column), resolved WDRG systems (fourth column) and with respect to the total population of white dwarfs (fourth column). The percentages obtained are quite similar for both models. In particular, for Model 29, nearly 23 per cent of double degenerate systems contain at least one He-core white dwarf and for WDMS systems around 14 per cent have a He-core white dwarf. Although the fraction of He-core is relatively important among these two sub-populations, the contribution to the entire population of white dwarfs (single and binary systems) is only around 1 per cent. This value is similar to the $\sim$0.5\% percentage of low-mass ($<$0.5\,M$_{\odot}$) white dwarfs among single white dwarfs estimated by \citet{Rebassa-Mansergas2011}.

Finally, we can compare the percentages obtained from our simulated sample with those derived from the observed {\it Gaia} sample. In Figure \ref{f:hewd} we plot the Hertzsprung-Russell diagram for the WDMS (red dots) and DWD (blue dots) systems considered in this work. As it is well known the location within the Hertzsprung-Russell diagram of a white dwarf depends on its mass and cooling time. For illustrative purposes we show the cooling tracks \citep[see][and references therein]{Camisassa2019} corresponding to masses from 0.5\,M$_{\odot}$ to 1.3\,M$_{\odot}$ in steps of 0.2\,M$_{\odot}$ for white dwarfs of hydrogen-pure (solid black line) and hydrogen-deficient (dashed black line) atmosphere models. White dwarfs with masses below 0.5\,M$_{\odot}$ can be safely considered helium-core white dwarf candidates \citep[e.g.]{Althaus2010}. We have marked  in Fig. \ref{f:hewd} as open symbols those objects with presumably helium-cores, founding 119 WDMS and 35 DWD of these systems, which represent $15\%$ and $17\%$ of its respective sub-population. These percentages are in agreement with other studies of the {\it Gaia} white dwarf population -- see, for instance, \citep{Rebassa-Mansergas2021}, where 46 low-mass objects have been found out of 235 WDMS (19\%). Although these observed percentages are relatively large, they  are in perfect agreement  within Poissonian errors with those derived in our simulated samples of Table \ref{t:He-WD}. It is beyond the scope of the present work to fully analyze the origin of these objects which is left for a forthcoming study.

\begin{table}
  \begin{tabular}{| c | c | c | c | c |}
  \hline
  Best fit & DWD$_{\rm res,\,He}$ & WDMS$_{\rm res,\,He}$ & WDRG$_{\rm res,\,He}$  & WD$_{\rm total,\,He}$\\ \hline

Model 29 & 22.66 & 14.26 &  65.19 &	1.11\\
Model 30 & 21.76 & 13.87 &  46.00 &	1.48 \\

\hline
  \end{tabular}
  
\caption{Percentages of He-core white dwarfs with respect to the different sub-populations (columns second to fourth) and with respect to the entire white dwarf population.}
\label{t:He-WD}
\end{table}

\subsubsection{The final orbital separation distribution}
\label{s:imf}

\begin{figure}
\includegraphics[width=1.2\columnwidth,trim=40 80 0 60, clip]{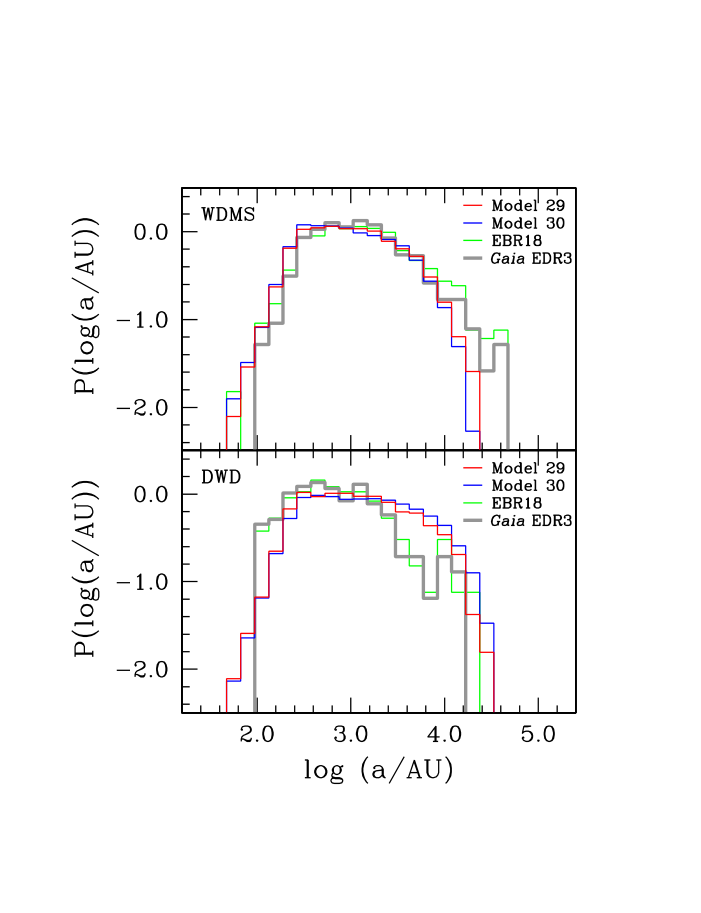}
\caption{Probability distribution of the final orbital separation of the WDMS (top panel) and DWD (bottom) samples. Shown as a gray line is the observed distribution from EDR3, while red and blue distributions correspond to synthetic Models 29 and 30, respectively. For comparative purposes we also shown the distribution (green) from \citet{ElBadry2018}.}
\label{f:asep}
\end{figure}

Another feature that we can analyze is the final orbital separation distribution, $a$, for resolved DWD and WDMS systems. In Figure \ref{f:asep} we show the distribution of $s$ for WDMS (top panel) and DWD (bottom panel) obtained from our observed EDR3 sample (gray histogram) and that presented in \cite{ElBadry2018} (EBR18; green histogram) for objects closer than 100 pc with those synthetically derived from our best fit Model 29 (red histogram) and Model 30 (blue histogram). Several interesting issues can be pointed out. First, no discrepancies are detected among the observed distributions. Second, when comparing them with the simulated sample we can state that, ignoring slight discrepancies in the tail distributions where statistical fluctuations are larger,  the simulated distribution for WDMS objects reasonably agrees with the observed one. However, the observed DWD distribution seems to decay faster than simulated predictions in the range $3.4<\log(a/\rm{AU})<4.2$.

In their study, \cite{ElBadry2018} thoroughly analyzed the observed separation of WDMS, DWD and main-sequence binaries. They reached the conclusion that the observed faster decay of the WDMS and DWD distributions occurring at $a\approx\! 3,000\,$AU and $a\approx\! 1,500\,$AU, respectively, was a consequence of asymmetrically mass loss during AGB phase which induced a small 'kick' during the formation of the white dwarf. Thus, many systems with initial relatively large separations can be unbounded, disappearing then from the resolved distribution of bound systems. This effect would explain then the steeper slope for wide WDMS and DWD systems. 

However, in our analysis we observe that there is no need to include such an effect for the WDMS distribution, where the simulated sample reasonably matches the observed one. Even the simulated sample appears to decay faster than the observed one for very large separations, $\log(a/\rm{AU})\gappr 3.8$. But, that is not the case for the DWD distribution, where clearly a faster decay occurs in the observed sample. These facts lead us to conclude that the effect of the white dwarf recoil should be weaker than that proposed in \cite{ElBadry2018}, thus having only a significant effect when two episodes of white dwarf formation occurs as in DWDs, while for WDMS systems the effect would be only to widening the orbits but not disrupting the systems. Consequently, asymmetric mass loss should be less efficient to induce a white dwarf 'kick' than previously proposed.

\subsection{The white dwarf mass distribution}
\label{s:mas}

\begin{figure}
\includegraphics[width=1.2\columnwidth,trim=47 150 0 70, clip]{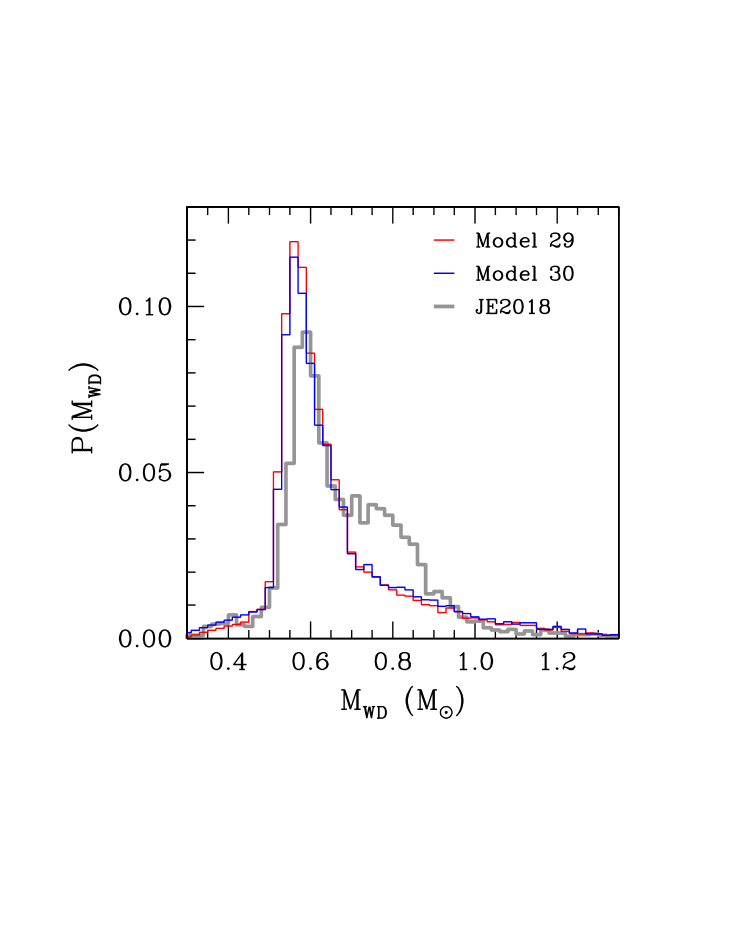}
\caption{Mass distribution of synthetic Model 29 (red line) and Model 30 (blue line) compared to the observed distribution of \citet{Jimenez-Esteban18} (gray line). Even that simulated models predict a fraction of mergers of $9\sim16\%$ is not enough to form the observed bump around $\sim0.8\,$\Msun.}
\label{f:mass}
\end{figure}

Finally we focus on how our best models compare with the observed white dwarf mass distribution. Several studies have been performed during the last decades devoted to obtain an unbiased mass distribution \citep[e.g.][]{Bergeron1992,Kepler2006,Rebassa-Mansergas2015,Rebassa-Mansergas2015b, Tremblay2016, McCleery2020}. The recent availability of {\it Gaia} data seems to indicate that, apart from a clear peak around $0.6\,$\Msun, the distribution presents a bump around $0.8\,$\Msun \citep{Jimenez-Esteban18,Kilic2020}. White dwarf mergers have been initially suggested as the source of this excess of massive objects \citep{Liebert2005, Kilic2018}, although several other hypotheses have been introduced such as the effect of a different slope in the initial-to-final mass relationship \citep{ElBadry2018a} or the existence of extra cooling delays \citep{Kilic2020}.

In Fig.\,\ref{f:mass} we display the white dwarf mass distribution for our best Model 29 (red line), Model 30 (blue line), and for comparative purposes the {\it Gaia} DR2 derived mass distribution (gray line) from \citet{Jimenez-Esteban18}. Our best fit model presents a clear peak at a slightly smaller value, $\sim0.56\,$\Msun, than the observed one,$\sim0.59\,$\Msun. It also displays an extended tail up to larger masses and it nicely recovers the fraction of low mass stars. However, no signs of a bump at $0.8\,$\Msun is detected in the simulated samples. The percentage of mergers of our best fit Model 29 is between $8\sim16$ per cent depending on the common-envelope phase treatment. In accordance with \cite{Temmink2020}, this percentage is clearly not enough to produce an excesses of massive objects. 

\subsection{The resolved binary fractions as a function of the sample volume}
\label{s:dis}

\begin{figure}
\includegraphics[width=1.2\columnwidth,trim=55 140 0 100, clip]{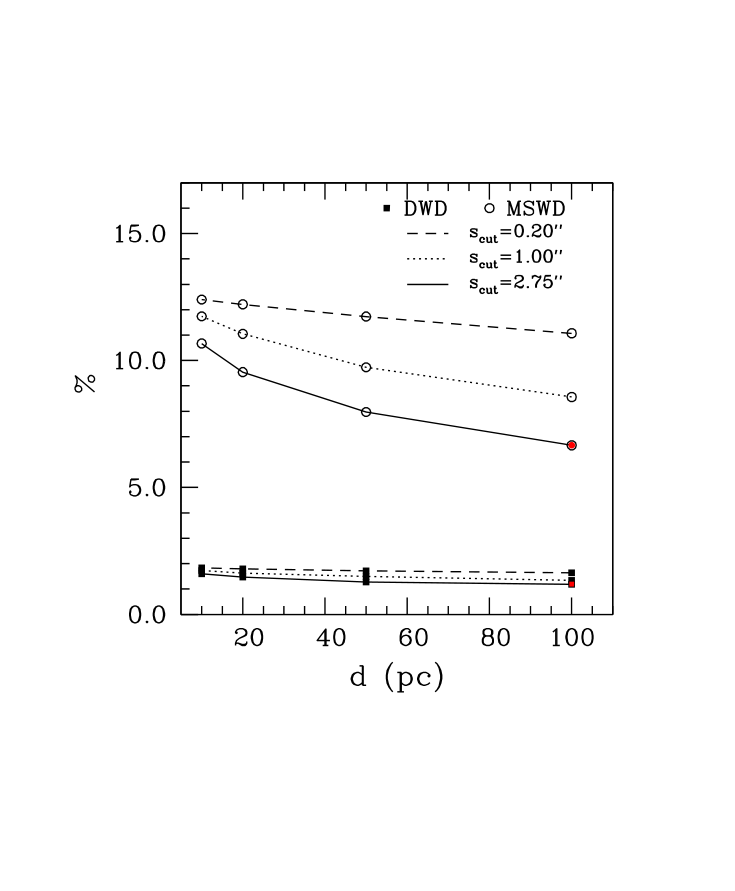}
\caption{Best-fit model percentages of resolved WDMS (open circles) and DWD (solid squares) as a function of the distance for different angular separation cuts of $s_{\rm cut}=0.2, 1.0$ and $2.75\arcsec$ (dashed, dotted and continuous line, respectively). Marked in red are the values obtained in this work.}
\label{f:discut}
\end{figure}

As previously stated in Section \ref{ss:como} the fraction of the different sub-populations can vary as a function of the observed volume. In particular, it is logically expected that the fraction of resolved systems increases as the observed sample is reduced. At the same time, the minimum angular separation adopted ultimately determines  the classification of a binary system as resolved or unresolved. The analysis of such effects becomes a hard task in any observed volume sample due to incompleteness biases and other selection effects. We take advantage of our synthetic population sample, in particular our best fit model, to analyze these issues.

In Figure \ref{f:discut} we show the percentages obtained from our best-fit model for resolved WDMS (open circles) and DWD (solid squares) as a function of the distance for different angular separation cuts of $s_{\rm cut}=0.2, 1.0$ and $2.75\arcsec$ (dashed, dotted and continuous line, respectively). We also show, marked in red, the percentages obtained by our best-fit model (1.19 and 6.66\%, for resolved DWD and WDMS, respectively) which perfectly matches the observed 100 pc {\it Gaia} sample for an angular separation cut of $s_{\rm cut}=2.75\arcsec$. We observe that, as the volume sample is reduced, the percentages of resolved objects increase as expected. In the case of resolved DWD, we obtain 1.5\% for the 20 pc sample, and even we can reach  1.8\% for a  $s_{\rm cut}=0.2\arcsec$. While, for resolved WDMS, the increment in the percentage is more significant as we reduce the volume sample. We obtain 9.54\% of resolved WDMS for a 20 pc sample and a 
$s_{\rm cut}=2.75\arcsec$, but it can be increased up to 12.5\% for a lower $s_{\rm cut}=0.2\arcsec$.

Percentages thus found by our best-fit model are in good agreement with those presented in \cite{Toonen2017} for the 20 pc sample (1.7 and 15\% for resolved DWD and WDMS, respectively). The analysis of more specific details of how observed samples have been constructed are beyond the objectives of the present paper and, in any case, it does not distort that our best-fit model presents a robust solution compatible with different observed volume samples.

\section{Conclusions}
\label{s:conclusions}

We have conducted a holistic analysis of the single and binary population of white dwarfs with the aim to find the parameters that best fit the observed fractions of the different sub-populations, with special emphasis on the resolved WDMS and DWD percentages. In order to carry out this study, we first have extracted from {\it Gaia}-EDR3 a nearly complete sample of white dwarfs identifying sources with accurate values of astrometric and photometric solutions. A comoving pairs identification method based on the tangential velocity for all pair candidates has allowed us to precisely determine the observed fraction of resolved DWD ($1.18\pm 0.10\%$) and WDMS ($6.31\pm 0.23\%$) with respect to the total population of white dwarf systems. Second, with the aid of a detailed population synthesis code, widely used in the modeling of single and binary white dwarf population, we thoroughly analyzed the space parameters.

Our first analysis indicates that the ratio of resolved WDMS to DWD systems mainly depends on the IMRD adopted, being the $n(q)\propto q^{-1}$ model, the one that better replicates the observed fractions. Moreover, the use of a BPL for the initial separation of the binary system can reasonably reproduce the observed fraction of resolved DWD and WDMS but at cost of generating an unrealistic percentage of mergers. Adopting a more sophisticated IMRD (MDS model) does not achieve as good results as the $q\propto q^{-1}$ model. Finally, the best models found so far have been achieved for a low binary fraction.

On the other hand, we can disregard an increasing IMRD, as well as we can ensure that the effects of the IMF slope or the common envelope treatment are of minor order with respect to the percentage of resolved systems.

On a second stage, a fine-tuning analysis reveals that the best fit can be achieved for an initial separation of
$f(a)\propto a^{-1}$, an IMRD  $n(q)\propto q^{n_q}$, with $n_q=-1.13^{+0.12}_{-0.10}$ and a binary fraction of $f_{\rm b}=0.32\pm 0.02$. 
Based on our distance to the observed percentages estimator (Eq. \ref{eq:estim}), we conclude that the previously best-fit model perfectly agrees with the observed fractions of the resolved sub-populations within a $1\sigma$ confidence level. The fraction of white dwarf mergers generated by this model --
$9\sim16\%$ depending on the common-envelope formalism adopted -- is also in accordance with published estimates \citep{Temmink2020}. Moreover, the binarity fraction of our best-fit, although lower than previous estimates, is in agreement at the $2\sigma$ confidence level (27\%-39\%) with the recent estimates from {\it Gaia}-EDR3 and LAMOST data release 5 \citep{Niu2021}. However, a broken power law with a twin excess for the IMRD model \citep{Moe2017,ELBadry2019} is only marginally compatible at a $6\sigma$ confidence level with the observed fraction of resolved sub-populations.

As as sub-product of our best-fit modeling we analyzed the following issues. First, the expected percentage of He-core stars is moderately low among the whole population, representing only $\sim 1\%$, although within the DWD and WDMS sub-populations it represents $\approx23\%$ and $\approx14\%$, respectively. Second, the percentage of mergers, as previously stated, is in the range of $9\sim16\%$ of the total population, which is not enough for explaining the observed excess of massive objects ($M\sim0.8$M$_{\odot}$) in the mass distribution. With respect to the final sky separation distribution, our analysis reveals that, although resolved DWDs present a faster decay for separations larger than $s\approx6,300\,$AU, no such a sign is observed in the WDMS distribution. These facts indicate that the proposed kick during the white dwarf formation due to asymmetric mass loss \citep{ElBadry2018} should be less effective. Finally, the analysis of the fraction of resolved systems as a function of the sample volume and the angular separation cut reveals that our best fit model is compatible with the observed results obtained for the 20 pc sample by \cite{Toonen2017}.

The excellent data released by {\it Gaia} has, thus, provided a precise benchmark for testing the suitability of population synthesis modeling of the white dwarf population. In our work, we have achieved a global solution of the space parameters compatible with the observed data. Although a more detailed analysis of some of the parameters is required, the solution proposed in this work can be used as a starting point in the modelling of the single and binary populations of white dwarfs in our solar neighbourhood.

\section*{Acknowledgements}

We acknowledge our anonymous referee for his/her positive comments and the resulting improvement of the paper. This work was partially supported by the MINECO grant  PID2020-117252GB-I00. ARM acknowledges support from grant RYC-2016-20254 funded by MCIN/AEI/10.13039/501100011033 and by ESF Investing in your future. This work has made use of data from the European Space Agency (ESA) mission {\it Gaia} (\url{https://www.cosmos.esa.int/gaia}), processed by the {\it Gaia} Data Processing and Analysis Consortium (DPAC, \url{https://www.cosmos.esa.int/web/gaia/dpac/consortium}). Funding for the DPAC has been provided by national institutions, in particular the institutions participating in the {\it Gaia} Multilateral Agreement. F.J.E. acknowledges financial support from the Spanish MINECO/FEDER through the grant AYA2017-84089 and MDM-2017-0737 at Centro de Astrobiolog\'ia (CSIC-INTA), Unidad de Excelencia Mar\'ia de Maeztu, and from the European Union's Horizon 2020 research and innovation programme under Grant Agreement no. 824064 through the ESCAPE - The European Science Cluster of Astronomy \& Particle Physics ESFRI Research Infrastructures project. This research has made use of the Spanish Virtual Observatory (\url{http://svo.cab.inta-csic.es}) supported from Ministerio de Ciencia e Innovaci\'on through grant PID2020-112949GB-I00. This research has made use of the SIMBAD database, operated at CDS, Strasbourg, France.

\section*{Data Availability Statement}
The data underlying this article are available in the article.  Supplementary material will be shared on reasonable request to the corresponding author.

\bibliographystyle{mnras} 
\bibliography{bin100pc} 



\bsp	
\label{lastpage}
\end{document}